\documentclass[letterpaper, 10 pt, conference, compress]{ieeeconf}  % Comment this line out if you need a4paper

\IEEEoverridecommandlockouts                              % This command is only needed if 
\usepackage{graphics} % for pdf, bitmapped graphics files
\usepackage{epsfig} % for postscript graphics files
\usepackage{mathptmx} % assumes new font selection scheme installed
\usepackage{times} % assumes new font selection scheme installed
\usepackage{amsmath} % assumes amsmath package installed
\usepackage{amssymb}  % assumes amsmath package installed
\usepackage{enumerate}
\usepackage{url}

\usepackage{soul,color}
\usepackage{nccmath,textcomp}
\usepackage{newtxmath}
\usepackage{makecell}
\usepackage{adjustbox}
\usepackage{comment}
\usepackage{booktabs}
\usepackage{tabularx}
\usepackage{array}
\usepackage{makecell} % 用于 \thead 命令
\usepackage{caption} % 用于修改表格标题格式

\usepackage{graphicx, caption, subcaption}
\usepackage[utf8]{inputenc}
\usepackage[T1]{fontenc}
\usepackage{cite}
\usepackage{hyperref}
\usepackage[ruled,noline,linesnumbered,noend]{algorithm2e}
\usepackage{multirow} % 用于跨行操作
\usepackage{microtype}

\usepackage{xcolor}

\title{\LARGE \bf
Optimizing LLM Inference Throughput via Memory-aware and SLA-constrained Dynamic Batching}

\author{Bowen Pang, Kai Li, Feifan Wang % <-this % stops a space
\thanks{Bowen Pang and Kai Li are with Noah's Ark Lab, Huawei Technologies, Beijing, China. {\tt\small pzkaixin@foxmail.com}, {\tt\small kaili.uest@gmail.com}}
\thanks{Feifan Wang is with the Department of Industrial Engineering, Tsinghua University, Beijing, China. {\tt\small wangfeifan@tsinghua.edu.cn}}
}

\begin{document}

\maketitle
\thispagestyle{empty}
\pagestyle{empty}

%%%%%%%%%%%%%%%%%%%%%%%%%%%%%%%%%%%%%%%%%%%%%%%%%%%%%%%%%%%%%%%%%%%%%%%%%%%%%%%%
\begin{abstract}
The increasing adoption of large language models (LLMs) necessitates inference serving systems that can deliver both high throughput and low latency. Deploying LLMs with hundreds of billions of parameters on memory-constrained GPUs exposes significant limitations in static batching methods. Current inference serving systems often treat batch sizes as fixed hyper-parameters, hindering real-time adaptation to varying system conditions. In this paper, we propose a dynamic batching method that continuously monitors memory utilization and adheres to service-level agreements (SLAs) to enable real-time batch size configuration adjustment. The method comprises two core components: a memory-aware batch scheduler that dynamically allocates GPU resources and a latency feedback mechanism that optimizes decoding processes under SLA constraints. The numerical experiments demonstrate throughput gains of 8\% to 28\% and capacity improvements of 22\% compared to traditional static batching methods, while maintaining full compatibility with existing inference infrastructure. These results highlight the effectiveness of dynamic batching in balancing computational efficiency and quality-of-service requirements for contemporary LLM deployment scenarios. The source code of this work is publicly available at \url{https://github.com/KevinLee1110/dynamic-batching}.
\end{abstract}

%%%%%%%%%%%%%%%%%%%%%%%%%%%%%%%%%%%%%%%%%%%%%%%%%%%%%%%%%%%%%%%%%%%%%%%%%%%%%%%%

\section{Introduction}

The widespread deployment of large language models (LLMs) \cite{achiam2023gpt,jaech2024openai,touvron2023llama,dubey2024llama,zeng2021pangu,liu2024deepseek,guo2025deepseek} in chatbots \cite{chatgpt}, code assistants \cite{githubcopilot}, and searching service \cite{microsoftcopilot} has created unprecedented demands for efficient inference serving systems that simultaneously achieve high throughput and low latency. In recent years, numerous optimization techniques have been developed, attempting to meet these demands. Strategies such as speculative decoding \cite{leviathan2023fast,chen2023accelerating}, kernel fusion \cite{dao2022flashattention}, and key-value (KV) cache management \cite{kwon2023efficient} have been proposed, alongside methods in request and iteration scheduling \cite{pang2025hybrid,jaillet2025online,wu2023fast} and distributed strategies \cite{zhong2024distserve,patel2024splitwise}. From the perspective of batching, a key configuration for enhancing throughput, current research includes continuous batching \cite{yu2022orca,agrawal2024sarathi} and PagedAttention \cite{kwon2023efficient}, both of which aim to increase batch sizes to enhance parallelism and throughput. However, all these traditional methods apply static batching policy and treat batch size as a hyper-parameter, which show limitations in handling dynamic workloads characterized by variable request patterns, sequence lengths, and latency requirements.

% A fundamental challenge emerges when serving LLMs with hundreds of billion parameters efficiently on memory-constrained GPUs - a reality facing most industrial deployments. 
% % While continuous batching \cite{yu2022orca} and PagedAttention \cite{kwon2023efficient} could be employed to economically manage GPU memory, this issue is still challenging, as the memory usage could grow and shrink dynamically.
% Current systems address this through request batching, which groups multiple user queries into single computational units to improve hardware efficiency. 

The primary limitation arises from rigid batch size management associated with static batching. Static batching methods pre-allocate fixed resources, risking either GPU underutilization in low-demand periods or memory overflow during traffic surges. Even the state-of-the-art dynamic inference serving systems focus primarily on token-level scheduling rather than proactive batch size optimization \cite{pang2025hybrid, jaillet2025online}. It creates two critical bottlenecks. First, it results in the conflict between memory efficiency and computational throughput. Large batch size increases parallelism, which further improves computational throughput, but also causes memory usage to grow linearly with sequence length. Second, it leads to the tension between computational throughput and quality-of-service requirements. Excessive batching risks maintaining latency compliance, directly impacting service quality.

\begin{figure}[t]
    \centering
    \includegraphics[width=0.98\columnwidth]{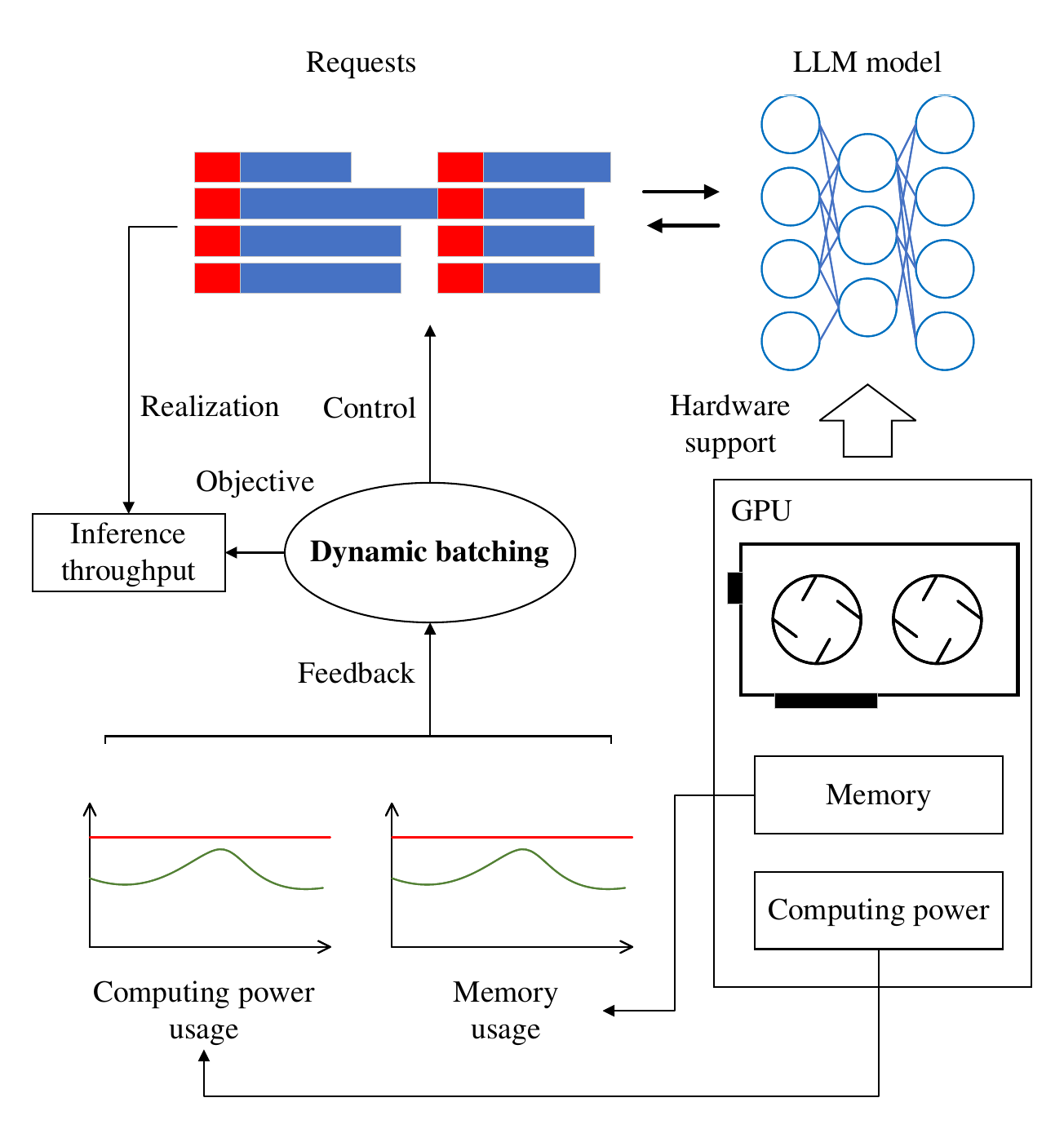}
%    \vspace{-0.1in}
    \caption{Dynamic batching as a real-time control problem}
%    \vspace{-0.2in}
    \label{dynamic_batching_structure}
\end{figure}

The real-world operational complexity intensifies these challenges. Diverse text generation tasks and application-specific latency constraints require batch schedulers simultaneously adapt to both hardware states and service objectives. Existing solutions often treat batch size as a fixed hyper-parameter while optimizing secondary factors, such as KV cache management. It leaves untapped potential in treating batch size as a first-class optimization decision variable for real-time dynamic batching. Developing an effective solution requires overcoming three key technical barriers: 1) The nonlinear relationship between batch size and system performance varies significantly across model architectures; 2) frequent batch adjustments may introduce computational overhead that offsets throughput gains; 3) coordinating competing objectives, e.g., preventing memory errors, maximizing throughput, and meeting latency targets, demands robust multi-criteria optimization under uncertainty. These challenges call for a fundamentally new approach that dynamically optimizes batch size configurations based on real-time system telemetry while maintaining operational stability. 

As illustrated in Fig. \ref{dynamic_batching_structure}, our work repositions static batch size hyper-parameter optimization as a real-time control problem that dynamically maximizes LLM inference throughput under GPU memory constraints with optional service-level agreements (SLAs) considerations for decoding latency. We first establish theoretical foundations and numerical relationships between batch size configurations and real-time throughput. Building on this analysis, we develop a memory-aware dynamic batching model that makes time-efficient batching decisions through continuous system monitoring. We further propose a variant that explicitly incorporates SLA constraints into the optimization framework. In particular, our method remains effective in prefill-decode (PD) fusion scenarios through adaptive chunk size determination. Numerical experiments demonstrate $8\%$ to $28\%$ throughput improvement over traditional static batching methods in vLLM implementations and $22\%$ capacity improvement under the SLA constraints.

Our key contributions are threefold. First, we introduce a dynamic batching method that dynamically adjusts batch size configurations based on instantaneous system states, effectively overcoming the limitations of static batching methods in both SLA-constrained and unconstrained environments. Second, we develop a rigorous mathematical model that characterizes the complex interplay between throughput maximization, memory constraints, SLA constraints, and dynamic batching, with theoretical guarantees. Third, comprehensive empirical evaluations reveal consistent throughput improvements across diverse operational scenarios, demonstrating our method's effectiveness in balancing computational efficiency with quality-of-service requirements. These advances collectively provide both theoretical insights and practical tools for optimizing LLM inference serving systems.

The remainder of this paper is organized as follows. Section \ref{problem_formulation} defines the dynamic batching problem and presents its mathematical formulation. Section \ref{solution and experiments} details our theoretical analysis, algorithmic implementations, and experimental validation. Section \ref{conclusion} discusses implications, limitations, and future research directions.

\section{Problem Formulation}
\label{problem_formulation}
\subsection{Problem description}
%LLM Inference Workflow and Batching

% System Overview (1 paragraph)

% Describe a typical LLM inference serving pipeline:
% Request arrival → Batching → Token generation (autoregressive decoding).
% Highlight GPU memory consumption drivers: model weights, KV caches, and batch size.
% Define key metrics: throughput (tokens/sec), latency (time-to-first-token, E2E), memory usage.

% Purpose: Define the technical scope.

% Key Points:

% LLM inference involves autoregressive token generation (prefill + decode phases).

% Batching groups multiple requests to parallelize computation.

% Memory consumption grows with batch size (activations, KV caches).

Batch size configuration critically impacts three interdependent metrics in LLM inference serving systems: throughput, latency, and memory usage. Throughput, defined by tokens processed per second, increases with larger batch sizes due to parallel computation across multiple requests, analogous to batch processing in manufacturing systems. However, this gain exhibits diminishing returns as memory constraints tighten. Decoding latency, measured as time between tokens (TBT), increases with batch size due to the higher computational cost, caused by the enlarged matrix dimensions in the matrix multiplication operations required for larger batches. The memory constraint stems from the KV cache, a dynamic data structure storing intermediate attention states. The KV cache size scales linearly with batch size and sequence length, establishing a hard capacity limit for concurrent requests.

\begin{figure}[t]
    \centering
    \includegraphics[width=0.98\columnwidth]{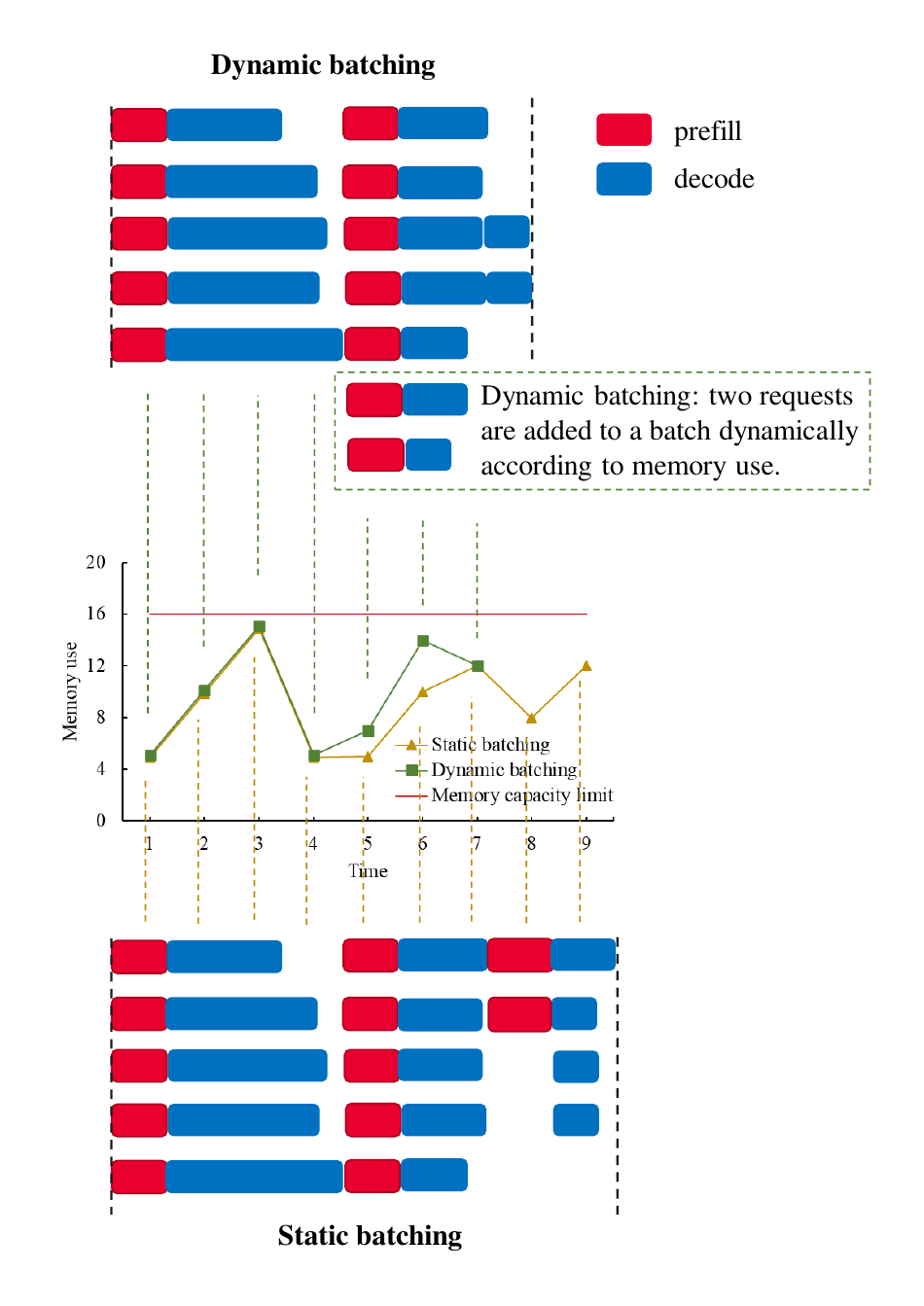}
%    \vspace{-0.2in}
    \caption{Dynamic batching according to memory use}
%   \vspace{-0.2in}
    \label{dynamic_vs_static}
\end{figure}

Current inference serving systems, such as vLLM, employ static batching, requiring operators to preset a fixed maximum batch size. This approach forces suboptimal trade-offs. Undersized batches waste GPU capacity through low utilization, while oversized batches risk memory overflows during demand spikes or long-sequence workloads. The static batching policy cannot adapt to time-varying request arrival patterns or heterogeneous sequence lengths, leading to either resource underutilization or SLA violations. This limitation persists despite advances in memory management techniques, such as PagedAttention \cite{kwon2023efficient}, which optimize memory allocation but do not dynamically change batch size. In this work, we propose a dynamic batching method to better utilize GPU memory as illustrated in Fig. \ref{dynamic_vs_static}.

% Although large batch size introduces memory overflow, engineers has already find methods include swapping and recomputation to overcome temporary memory overflow \cite{kwon2023efficient}. Swapping method evicts exceeding GPU memory to CPU and temporarily save it when some prompts are preempted. These requests will be brought back later when GPU is free. Recomputation happens when these requests are bought back to recompute the KV-cache. These methods are effective but will raise higher computation load afterwards. Therefore, we can also treat the memory limit as a soft constraint and simply reduce the probability of exceeding memory limit.

% To address memory overflow issues associated with large batch sizes, techniques such as swapping and recomputation have been employed \cite{kwon2023efficient}. The swapping method involves temporarily moving data from GPU memory to CPU memory when capacity is exceeded, with the data being moved back to the GPU when space becomes available. Recomputation occurs when these requests are brought back to the GPU and the KV cache is recomputed. Although effective, these methods increase computational load. Therefore, an alternative approach treats memory limit as a soft constraint, focusing on minimizing the probability of exceeding the memory limit, rather than strictly preventing it.

Although large batch sizes face memory overflow risk, researchers have developed mitigation techniques, such as swapping and recomputation \cite{kwon2023efficient}. The swapping method involves temporarily moving data from GPU memory to CPU memory when capacity is exceeded. The data are moved back to the GPU when space becomes available. Recomputation occurs when regenerating the KV cache from intermediate states during request reactivation. While these methods effectively resolve temporary memory overflow issues, they incur non-negligible overhead. Swapping introduces data migration latency, and recomputation increases computational redundancy. However, it suggests that memory limitations could be treated as soft constraints with probabilistic guarantees, rather than absolute capacity boundaries requiring strict enforcement.

\subsection{Problem formulation}
% Modern LLM inference systems process request sequences through iterative token generation. Each request arrives with variable input lengths under two constraints: fixed GPU memory capacity and SLA-bound latency guarantees. Batched execution improves throughput but creates conflicting objectives. Maximizing batch sizes enhances hardware utilization while risking memory overflows and latency violations. The inference system requires a real-time batch size adaptation to simultaneously achieve throughput maximization, memory constraint satisfaction, and per-request latency compliance.

The core challenge lies in designing an adaptive batch scheduler that dynamically adjusts batch sizes under operational constraints. We formulate it as an online optimization problem, where the batch size at time $t\in \mathbb{Z}_+$ is denoted by $b_t \in \mathbb{Z}_+$, the maximum batch size limit is denoted by $B_{\text{max}} \in \mathbb{Z}_+$, and the minimum batch size limit is denoted by $B_{\text{min}} \in \mathbb{Z}_+$. The following constraints should be satisfied.  
\begin{itemize}
    \item Memory constraints: The memory consumption under batch size $b_t$ must satisfy $M\left(b_t\right) \leq M_{\max}$, where $M\left(b_t\right) \in \mathbb{R}_+$ refers to dynamic KV caches and $M_{\max} \in \mathbb{R}_+$ represents the remaining GPU memory after allocating space for LLM parameters and preallocating space for temporary activations. 
    %This approach is commonly utilized in practice, for instance, in the vLLM framework \cite{kwon2023efficient}.
    % \item Memory constraints: Total memory consumption $M(b_t) \leq M_{\max}$, where $M(b_t)$ combines LLM parameters, dynamic KV caches ($\propto b_t \cdot L_{\max}$), and temporary activations.
    \item Workload dynamics: Workload dynamics are associated with non-stationary request arrival rate at time $t$, denoted by $\lambda\left(t\right)\in \mathbb{R}_+$, and heterogeneous sequence length as a discrete random variable, denoted by $L$. Let $L_{\text{max}}\in \mathbb{Z}_+$ be the possibly largest sequence length. 
    %Thus, $L$ has probability distribution given by $\mathbb{P}\left(L = l\right) = p_l\in \left(0,1\right)$, for $l = 1, 2, \cdots, L_{\text{max}}$, and $\sum_{l=1}^{L_{\text{max}}} p_l = 1$.
    \item Latency requirements: Per-request decoding latency is a function of batch size $b_t$, denoted by $D\left(b_t\right)\in \mathbb{R}_+$. It is observed that $D(b_t)$ linearly depends on batch size $b_t$. We use $D_{\text{SLA}}\in \mathbb{R}_+$ to define the largest allowed latency according to SLA constraints. Thus, we have constraint $D(b_t) \leq D_{\text{SLA}}$.
\end{itemize}

The problem at each scheduling interval is formulated as an optimization model as follows.
\begin{align}
    \underset{b_t}{\text{max}} \quad  &\Phi\left(t\right) = \mathbb{E} \left[\frac{L \lambda\left(t\right)}{\tau_{\text{step}}\left(b_t\right) n\left(b_t\right)}\right], \label{eq:objective} \\
    \label{eq:memory_constraint} 
     \text{s.t.} \quad &\mathbb{P}\left(M\left(b_t\right) > M_{\max}\right) \leq \epsilon_\text{M},\\
    &D\left(b_t\right) - D_{\text{SLA}} \leq \epsilon_\text{D}, \label{eq:latency_constraint}  \\
    &b_t \in \mathbb{Z}_+, \nonumber
\end{align}
where $\tau_{\text{step}}\left(b_t\right) \in \mathbb{R}_+$ denotes the computation time per decoding step for batch size $b_t$, and $n\left(b_t\right) \in \mathbb{Z}_+$ denotes the number of inference steps required at this scheduling interval. In the objective function \eqref{eq:objective}, $\Phi\left(t\right) \in \mathbb{R}_+$ represents the expected token throughput, incorporating both batch size efficiency and sequence length distribution. Constraint \eqref{eq:memory_constraint} guarantees that the probability of the memory in use $M\left(b_t\right)$ exceeding the memory limit $M_{\max}$ should be lower than a threhold, denoted by $\epsilon_\text{M} \in \left(0,1\right)$. Constraint \eqref{eq:latency_constraint} enforces SLA satisfaction with an absolute error $\epsilon_\text{D} \in \mathbb{R}_+$.
%Challenges of Static Batching

% Static vs. Dynamic Batching Limitations (1-2 paragraphs)

% Static Batching: Fixed batch sizes lead to:
% Memory overflow during peak loads.
% Low GPU utilization during off-peak periods.
% Existing Dynamic Batching (e.g., continuous batching):
% Focus on incremental updates but lack proactive batch size tuning.
% Fail to balance memory, throughput, and latency holistically.

In this problem, memory consumption \( M\left(b_t\right) \) scales linearly with batch size \( b_t \) and sequence length $L$ due to KV cache overheads.
Decoding latency \( D\left(b_t\right) \) increases linearly with \( b_t \) due to GPU parallelism, but the relationship varies depending on hardware and LLM architectures. Bursty request arrivals, such as sudden spikes in \( \lambda\left(t\right)\), may force abrupt batch size reductions to prevent out of memory (OOM) errors, leading to throughput instability. Heterogeneous sequence lengths complicate memory prediction, as long sequences in a batch may exhaust memory even at small \( b_t \).

\section{Solution and Experiments}
\label{solution and experiments}
While some applications operate under strict SLA constraints that specify performance guarantees for TBT, others do not have explicit SLA constraints and instead prioritize maximizing throughput to process high volumes of requests efficiently. For instance, applications, such as real-time customer support systems and conversational AI used in healthcare or finance, typically enforce stringent SLA constraints to ensure low-latency responses and high reliability. On the other hand, batch-processing tasks, such as document summarization, offline content generation, and large-scale data labeling, emphasize throughput maximization without strict SLA constraints. Understanding the distinct requirements of these applications is essential for designing inference serving systems that can effectively balance latency, throughput, and computational resource utilization in heterogeneous deployment environments. In this section, we introduce two solution methods for scenarios with and without SLA constraint \eqref{eq:latency_constraint}, respectively. 

\subsection{Solution without SLA constraint}
We start with the objective function $\Phi(t)$, which is equivalent to
\begin{equation}
 \Phi\left(t\right)=\mathbb{E}\left[\frac{L}{n\left(b_t\right)}\right] \frac{\lambda\left(t\right)}{\tau_{\text{step}}\left(b_t\right)},
\end{equation}
as the randomness over output lengths is independent than both the arrivals of requests and the batching policy. We assume full batch utilization, which means that, at each inference round of decoding, the number of working requests equals the batch size $b_t$. Thus, we have

\begin{equation}
n\left(b_t\right)=\frac{L \lambda\left(t\right)}{b_t}.
\end{equation}
This assumption is naturally true when we seek to maximize throughput with relatively large arrival rate $\lambda\left(t\right)$. Under this scenario, 

\begin{align}
\Phi\left(t\right)=\frac{b_t}{\tau_{\text{step}}\left(b_t\right)}.
\end{align}

% Knowing that $\tau_{\text{step}}(b_t)$ is a positive linear function of $b_t$, we draw an illustration of $\Phi(t)$ in Fig. \ref{fig1} using blue line. From the curve we know that $\Phi(t)$ is increasing with $b_t$. It is easy to know that maximizing throughput without SLA constraint equals to maximizing $b_t$.

\begin{figure}[t]
    \centering
    \includegraphics[width=0.98\columnwidth]{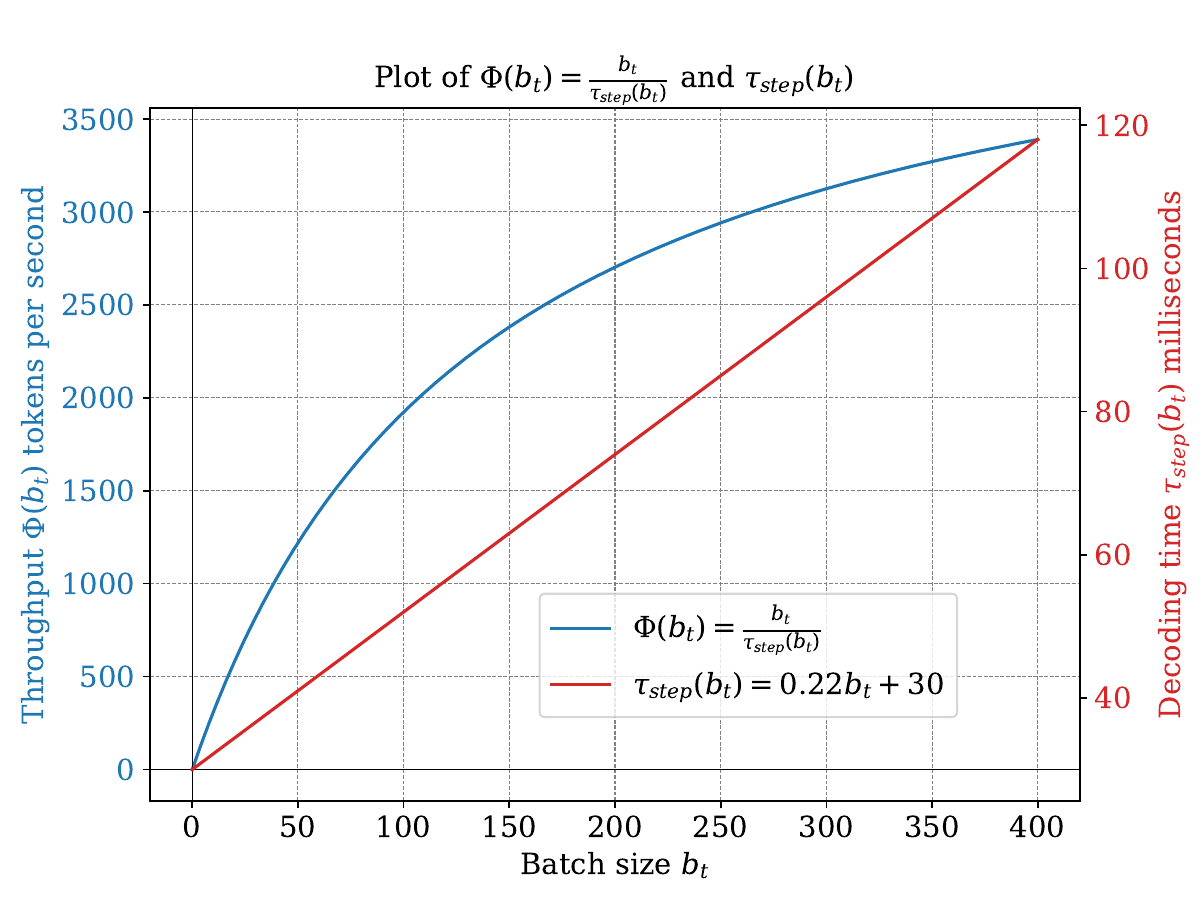}
%    \vspace{-0.1in}
    \caption{Relationship among dynamic batch size, inference throughput, and decoding time}
%   \vspace{-0.2in}
    \label{throughput_and_batchsize}
\end{figure}

The numerical relationship between system throughput $\Phi\left(t\right)$ and batch size $b_t$ is illustrated by the blue curve in Fig. \ref{throughput_and_batchsize}. Given the established positive linear dependence of $\tau_{\text{step}}\left( b_t \right)$ on $b_t$, the experimental results demonstrate that $\Phi\left(t\right)$ maintains a monotonic increasing relationship with $b_t$, while exhibiting a diminishing marginal gain. Specifically, the first derivative $d\Phi\left(t\right)/db_t$ remains positive throughout the operational domain, whereas the second derivative $d^2\Phi\left(t\right)/db^2_t$ becomes negative, indicating characteristic concave curvature. This concave progression implies that, although increasing $b_t$ effectively enhances throughput, the incremental benefit per unit $b_t$ decreases progressively. Consequently, the throughput maximization problem under no SLA constraints is reduces to maximize batch size $b_t$ in the feasible region $\{b_t \mid  M\left( b_t \right)\leq M_{\max} \}$.

% After each round of decoding, the increment of tokens is 
% \begin{align}
%     \lambda^{\text{token}} = b_t,
% \end{align}

% and the number of releasing token is 
% \begin{align}
%     \mu^{\text{token}}=\frac{b_t*\left(l_{\text{in}}+l_{\text{out}}\right)}{l_{\text{out}}}.
% \end{align}

% Therefore, if we assume there is a token limit $\eta$ for all requests is being processed in the serving system, we can calculate the following queueing length probability with arrival rate $\lambda^{\text{token}}$ and departure rate $\mu^{\text{token}}$.

% \begin{align}
%     P(L>\eta) = \left(\frac{\lambda^{\text{token}}}{\mu^{\text{token}}}\right)^{\eta}.
% \end{align}

To illustrate the relationship between $b_t$ and $M\left(b_t\right)$, we provide the following analysis. Suppose that the random variables representing the input and output token numbers on the $i$th batch, for $i=1,2,\cdots,b_t$, are denoted by $l_{\text{in},i}$ and $l_{\text{out},i}$, respectively. Let $\eta\in \mathbb{Z}_+$ be the maximum number of tokens available in the system given the memory limit  $M_{\max}$. The total number of tokens in the system at steady state is the sum of independent and identically distributed random variables, expressed as
\begin{equation}
    S = \sum_{i=1}^{b_t} \left(l_{\text{in},i} + l_{\text{out},i}\right).
\end{equation}
For any $i$, the expected value of $S$, denoted by $\mu_S$, is given by
\begin{equation}
\label{mu_s}
    \mu_S = b_t \left(\mathbb{E}[l_{\text{in},i}] + \mathbb{E}[l_{\text{out},i}]\right).
\end{equation}
The variance of $S$, denoted by $\sigma_S^2$, is given by
\begin{align}
\label{sigma_s}
    \sigma_S^2 = b_t \left(\text{Var}\left(l_{\text{in},i}\right) + \text{Var}\left(l_{\text{out},i}\right)\right).
\end{align}
Given that the batch size $b_t$ is typically large, we can invoke the Central Limit Theorem to approximate $S$ as normally distributed, i.e., $S \sim \mathcal{N}\left(\mu_S, \sigma_S^2\right)$. Consequently, we can use the cumulative distribution function of the standard normal distribution, denoted by $\Theta\left(\cdot \right)$, to calculate the following probability.

\begin{equation}
\label{probability_theta}
    \mathbb{P}\left(S > \eta \right) \approx 1 - \Theta\left(\frac{\eta - \mu_S}{\sigma_S}\right).
\end{equation}
Thus, 
\begin{align}
    \notag\mathbb{P}\left(M\left(b_t\right) > M_{\max}\right) &= \mathbb{P}\left(S > \eta\right) \\
    \label{probability_cal}
    & \approx 1 - \Theta\left(\frac{\eta - \mu_S}{\sigma_S}\right) \leq \epsilon_\text{M}.
\end{align}
Given that $b_t$ is given by \eqref{mu_s}, \eqref{sigma_s}, and \eqref{probability_cal}, let $\theta=\Theta^{-1}\left(1-\epsilon_\text{M}\right)$, and we have

\begin{align}
\label{eq:b_tleq}
     b_t \leq \left( \frac{ \sqrt{\left(\sigma_S \theta \right)^2  + 4\mu_S \eta} -\sigma_S \theta}{2\mu_S} \right)^2.
\end{align}
According to equations \eqref{mu_s}, \eqref{sigma_s}, \eqref{probability_cal}, and $\theta = \Theta^{-1}\left(1 - \epsilon_\text{M}\right)$, the batch size $b_t$ can be obtained as follows.

\begin{align}
    & \eta - b_t \left(\mathbb{E}\left[l_{\text{in},i}\right] + \mathbb{E}\left[l_{\text{out},i}\right]\right) \geq L_0,\\
    & b_t \leq \frac{\eta-L_0}{\mathbb{E}\left[l_{\text{in},i}\right] + \mathbb{E}\left[l_{\text{out},i}\right]},
\end{align}
where $L_0=\eta-\left(\theta\sigma_S+\mu_S\right)$ can be computed offline and updated online periodically using equation \eqref{probability_theta}. This $L_0$ can be considered as a safety buffer allocated to prevent from hitting memory limit. This method allows for real-time calculation of the batch size $b_t$ as a linear function with limited online information, including only current average input and output length $\left(\mathbb{E}\left[l_{\text{in},i}\right] + \mathbb{E}\left[l_{\text{out},i}\right]\right)$. The corresponding algorithm is detailed in Algorithm \ref{alg:memory_based_dynamic_batching}.

\begin{algorithm}
\caption{Memory constrained dynamic batching}\label{alg:memory_based_dynamic_batching}
\DontPrintSemicolon
\KwIn{Previous batch size $b_{t-1}$, expected prompt length $\mathbb{E}\left[l_{\text{in},i}\right]$ and output length $\mathbb{E}\left[l_{\text{out},i}\right]$}
\KwOut{Current batch size $b_{t}$}
\SetKwFunction{Fdb}{BatchingMemory}
\SetKwProg{Fn}{Function}{:}{}
 $L_0 \gets \eta-(\theta\sigma_S+\mu_S)$\;
\Fn{\Fdb{$b_{t-1}, \mathbb{E}\left[l_{\mathrm{in},i}\right], \mathbb{E}\left[l_{\mathrm{out},i}\right]$}}{
    $b_{t} \gets b_{t-1}$\;
    \If{$N_{t-1}^\mathrm{d} > 0 \wedge N_{t-1}^\mathrm{p} > 0$}{
        $b_{t} \gets \left \lfloor \frac{\eta-L_0}{\mathbb{E}[l_{\text{in},i}] + \mathbb{E}[l_{\text{out},i}]} \right \rfloor$\;
        $b_t \gets \min \{ \max \{ b_t, N_{t-1}^\text{d} \}, B_{\text{max}} \}$\;
    }
\KwRet $b_{t}$\;
}
\end{algorithm}

In the algorithm, we compute the number of pre-allocated tokens periodically. Subsequently, we introduce a function, \texttt{BatchingMemory}, designed to dynamically modify the current batch size \( b_{t} \). Within this function, \( b_t \) is initialized to the previous batch size \( b_{t-1} \) by default. Let $N_{t}^\text{p}\in \mathbb{Z}_+$ and $N_{t}^\text{d}\in \mathbb{Z}_+$, for $t\in \mathbb{Z}_+$, be the numbers of prefill requests and decode requests, respectively.  Only if both \( N_{t-1}^\text{p} \) and \( N_{t-1}^\text{d} \) exceed zero, the current batch size can be adjusted in accordance with equation~(\ref{eq:b_tleq}). Here, the condition \( N_{t-1}^\text{d} > 0 \) ensures that \( M\left(b_t\right) > 0 \), while \( N_{t-1}^\text{p} > 0 \) implies that no adjustment to the batch size is necessary in the absence of prefill requests. Line 6 in the algorithm guarantees that \( b_t \) must exceed the number of currently running requests and remain below the maximum batch size limit \( B_{\max} \). In practical inference serving systems such as vLLM, our algorithm can be implemented using blocks rather than relying on the number of tokens.

% \kai{$M_{\max}$ and $M(t)$ may contain memory for model parameters. Temporary activations are only used in time $t$. I need a variable for representing pure KV cache.}

\subsection{Solution with SLA constraint}
According to the result in Fig. \ref{throughput_and_batchsize}, scenarios without SLA constraints can be simplified to a problem that maximizes $b_t$ under the constraints of memory. Let $b_t^{\text{mem}}\in \mathbb{Z}_+$ be such maximized $b_t$ only under the constraints of memory. Then, we use $b_t^{\text{SLA}}\in \mathbb{Z}_+$ to denote the maximized $b_t$ only under SLA constraint. We can calculate $b_t^{\text{mem}}$ using Algorithm \ref{alg:memory_based_dynamic_batching}. The algorithm used to obtain $b_t^{\text{SLA}}$ is provided in Algorithm \ref{alg:sla_based_dynamic_batching}. Thus, the optimal batch size can be obtained by $b_t^* = \min\{b_t^{\text{mem}},b_t^{\text{SLA}}\}$. The information provided in Fig. \ref{throughput_and_batchsize} also offers a straightforward method to explore how SLA, batch size, and throughput influence each other. For example, if an SLA with a decoding time of 50 milliseconds is specified, the batch size $b_t$ can be estimated from the red line to be approximately 100. The throughput $\Phi\left(b_t\right)$ can then be estimated from the blue line to be around 1,900 tokens per second. If the service provider allows the SLA to increase to 80 milliseconds, the batch size can increase to 230, and the throughput can rise to 2,700 tokens per second.

\begin{algorithm}
\caption{SLA constrained dynamic batching}\label{alg:sla_based_dynamic_batching}
\DontPrintSemicolon
\KwIn{$D_{\text{SLA}}$ and the search space $[B_{\mathrm{min}}, B_{\mathrm{max}}]$}
\KwOut{Current batch size $b_{t}$}
\SetKwFunction{Fdb}{BatchingSLA}
\SetKwProg{Fn}{Function}{:}{}
Initialize $b_{0}^{\mathrm{low}}\gets B_{\mathrm{min}}, b_{0}^{\mathrm{high}}\gets B_{\mathrm{max}}$\;
\Fn{\Fdb{$D_{\mathrm{SLA}}, B_{\mathrm{min}}, B_{\mathrm{max}}$}}{
    Get recent average decode latency $\bar{\tau}$\;
    Get recent average decode batch size $\bar{b}$\;
    \uIf{$\bar{\tau} > D_{\mathrm{SLA}} + \epsilon_\mathrm{D}$}{
        $b_{t}^{\mathrm{high}} \gets \max \{\bar{b}, b_{t-1}^{\mathrm{low}} + \alpha \}$\;
        $b_{t}^{\mathrm{low}} \gets \max \{ b_{t-1}^{\mathrm{low}} - \delta, B_{\text{min}} \}$\;
    }
    \uElseIf{$\bar{\tau} < D_{\mathrm{SLA}} - \epsilon_\mathrm{D}$}{
        $b_{t}^{\mathrm{low}} \gets \min \{ \bar{b}, b_{t-1}^{\mathrm{high}} - \alpha \}$\;
        $b_{t}^{\mathrm{high}} \gets \min \{ b_{t-1}^{\mathrm{high}} + \delta, B_{\text{max}} \}$\;
    }
    \Else{
        $b_{t}^{\mathrm{high}} \gets \min \{ \bar{b} + \lfloor \alpha/2 \rfloor, B_{\text{max}} \}$\;
        $b_{t}^{\mathrm{low}} \gets \max \{ \bar{b} - \lfloor \alpha/2 \rfloor, B_{\text{min}} \}$\;
    }
    $b_t \gets \lfloor (b_{t}^{\mathrm{low}} + b_{t}^{\mathrm{high}})/2 \rfloor$ \;
    $b_t \gets \min \{ \max \{ b_t, N^\text{d}_{t-1} \}, B_{\text{max}} \}$\;
\KwRet $b_t$\;
}
\end{algorithm}

In Algorithm~\ref{alg:sla_based_dynamic_batching}, we introduce an online SLA-constrained dynamic batching algorithm designed to adjust to variations in user request sizes. The hyper-parameters $D_{\text{SLA}}$, $B_{\min}$, and $B_{\max}$ are specified by users, representing the SLA for decoding time, the hard upper and lower bounds for batch size, respectively. We define $b_t^{\text{low}} \in \mathbb{Z}_+$ and $b_t^{\text{high}} \in \mathbb{Z}_+$ as the temporary lower and upper bounds of the algorithm's search space. Initially, $b_t^{\text{low}}$ and $b_t^{\text{high}}$ can be set equal to $B_{\min}$ and $B_{\max}$. The algorithm primarily employs a binary search technique. To reduce noise during the search, we incorporate a small constant $\delta \in \mathbb{Z}_+$ as a corrective element. Additionally, a constant $\alpha \in \mathbb{Z}_+$ is utilized to control the interval between $b_t^{\text{low}}$ and $b_t^{\text{high}}$. The algorithm demonstrates efficiency and robustness in handling online request arrivals.

% In Algorithm~\ref{alg:sla_based_dynamic_batching}, we present an online SLA-constrained dynamic batching algorithm that adapts to variations in user request sizes. $D_{\text{SLA}},B_{\min},B_{\max}$ are hyperparameters given by user, representing the SLA on decoding time, hard batch size upper bound and lower bound, respectively.
% We define $b_{\text{low}} \in \mathbb{Z}_+$ and $b_{\text{high}} \in \mathbb{Z}_+$ as the temporary lower bound and upper bound of the search space for the algorithm, respectively. The values of $b_{\text{low}}$ and $b_{\text{high}}$ can be initialized as equal to $B_{\min}$ and $B_{\max}$.
% The algorithm is fundamentally based on a binary search procedure. To mitigate noise during the search process, we incorporate a small constant $\delta \in \mathbb{Z}_+$ as a corrective measure. In addition, a constant $\alpha \in \mathbb{Z}_+$ is used to regulate the interval between $b_{\text{low}}$ and $b_{\text{high}}$. The algorithm effectively exhibits efficiency and robustness in managing online request arrivals.

\begin{table*}
    \centering
    \caption{Throughput using static vs. dynamic batching under different LLMs and prompts settings}
    \label{tab:numerical-noSLA}
    \begin{tabular}{r|ccc|cc|c}
    \toprule
    % settings / throughput(tokens per second)
        % LLM & input (tokens) & output (tokens) & request num & static batching & dynamic batching & \textbf{improvement}\\
        \multirow{2}{*}{\textbf{LLM}} & \multicolumn{3}{c|}{\textbf{Experimental Settings}} & \multicolumn{2}{c|}{\textbf{Throughput (token/s)}} &  \multirow{2}{*}{\textbf{Improvement}}\\
        & Prompt Tokens & Output Tokens & Request Num & Static Batching & Dynamic Batching & \\
    \midrule
        LLaMA-65B & 68.4 & 344.5 & 1319 & 1983 & 2146 & \textbf{8.2\%}\\
        LLaMA3-70B & 68.4 & 454.4 & 1319 & 3153 & 3357 & \textbf{6.5\%}\\
        LLaMA3-70B & 191.0 & 381.9 & 3000 & 2296 & 2575 & \textbf{12.2\%}\\
        PanGu-7B & 128 & 128 & 1000 & 2305 & 2956 & \textbf{28.2\%}\\
        PanGu-38B & 128 & 128 & 1000 & 2215 & 2569 & \textbf{26.0\%}\\
        PanGu-135B & 128 & 128 & 1000 & 1342 & 1449 & \textbf{8.0\%}\\
    \bottomrule
    \end{tabular}
\end{table*}

\begin{table*}
    \centering
    \caption{Throughput with SLA using static vs. dynamic batching under different LLMs and prompts settings}
    \label{tab:numerical-SLA}
    \begin{tabular}{r|cccc|cc|cc|c}
    \toprule
        % LLM & $D_{\text{SLA}}$ & input (tokens) & output (tokens) & request num & static batching capacity & static batching throughput & dynamic batching capacity &dynamic batching throughput & \textbf{improvement}\\
        \multirow{3}{*}{\textbf{LLM}} & \multicolumn{4}{c|}{\textbf{Experimental Settings}} & \multicolumn{2}{c|}{\textbf{Capacity (qps)}} & \multicolumn{2}{c|}{\textbf{Throughput (token/s)}} &  \multirow{3}{*}{\textbf{Improvement}}\\
        & $D_{\text{SLA}}$ & \makecell{Prompt\\ Tokens} & \makecell{Output\\ Tokens} & \makecell{Request\\ Num} & \makecell{Static\\ Batching} & \makecell{Dynamic\\ Batching} & \makecell{Static\\ Batching} & \makecell{Dynamic\\ Batching} &\\
    \midrule
        LLaMA-65B & 50ms & 237.7 & 416.2 & 3000 & 3 & 3.3 & 1190  &1223 & \textbf{2.7\%}\\
        LLaMA3-70B & 50ms & 256.6 & 61.5 & 3000 & 5.4 & 6.6 & 331 & 405 & \textbf{22.4\%}\\
        LLaMA3-70B & 50ms & 256.6 & 447.5 & 3000 & 3.0 & 3.8 & 1322 & 1665 & \textbf{25.9\%}\\
    \bottomrule
    \end{tabular}
\end{table*}

\subsection{Numerical experiments}
The numerical experiment results using real LLM and real prompts are presented in TABLE \ref{tab:numerical-noSLA} and TABLE \ref{tab:numerical-SLA}.

In TABLE \ref{tab:numerical-noSLA}, we present the results of a comparative analysis between the baseline, which employs a static batch size as configured by vLLM, and the throughput achieved using our proposed dynamic batching method. In this experiment, the request arrival rate is set to infinite, meaning that all requests are sent to the LLMs simultaneously at the start. This setup allows us to assess the maximum potential token generation rate. We have implemented our method directly on vLLM by modifying their scheduling architecture to accommodate dynamic batching. The results indicate that our method increases throughput by $8\%$ to $28\%$ across all scenarios, irrespective of different LLM configurations and prompts. We further examine the average GPU utilization rate, and it increases from less than 40\% to nearly 50\%.

TABLE \ref{tab:numerical-SLA} presents the results of the baseline method using vLLM with our method, which employs SLA-constrained dynamic batching. Under the constraints imposed by the SLA, we compare the maximum request sending rate achievable by both the baseline and our method. The results indicate that our method enhances the throughput of various LLMs and different prompts while adhering to the SLA constraints. Moreover, the third line is implemented with PD fusion scenario, showing that our method is also valid for determining chunk size.

\begin{figure}[t]
    \centering
    \includegraphics[width=0.98\columnwidth]{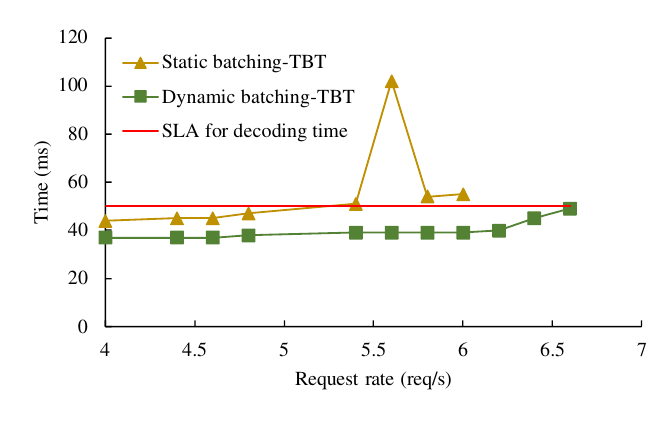}
%    \vspace{-0.2in}
    \caption{Capacity with SLA 50ms: dynamic vs. static batching}
%    \vspace{-0.2in}
    \label{capacity}
\end{figure}

In Fig. \ref{capacity}, we adopt the definition of capacity from \cite{agrawal2024sarathi}, which refers to the maximum number of requests that a system can handle while meeting the specified SLA targets. Higher capacity values indicate better performance. With an SLA on decoding time requirement of 50 milliseconds, our dynamic batching method increases the system's capacity from 5.4 to 6.6 queries per second (qps).

\section{Conclusion}
\label{conclusion}

In this paper, we aim at LLM inference optimization in scenarios with and without SLA constraints. Our main contribution is to develop a dynamic batching method that can outperform traditional static batching methods. Through theoretical analysis and numerical experiments, we systematically explore the interrelationships between throughput, dynamic batch size, memory limitations, and decoding latency. Specifically, we develop two novel algorithms. Algorithm \ref{alg:memory_based_dynamic_batching} introduces a memory-aware dynamic batching mechanism, while Algorithm \ref{alg:sla_based_dynamic_batching} extends this framework to incorporate SLA constraints. Extensive experiments using the widely adopted vLLM inference engine reveal that both algorithms achieve substantial throughput improvements of up to 28\% over static batching methods. Under SLA constraints, the dynamic batching method also enhances capacity by 22\%. Importantly, our implementation maintains compatibility with existing vLLM architectures, requiring minimal code modifications for practical deployment.

For future research, we identify three areas for exploration. First, the computational efficiency of Algorithm \ref{alg:memory_based_dynamic_batching} could be improved by replacing the current heuristic approach with the rigorous formulation in \eqref{eq:b_tleq}. 
%Second, while our SLA-aware method effectively manages decoding latency, further investigation is needed to assess its impact on time-to-first-token (TTFT) metrics, which are crucial for interactive applications. 
Second, due to hardware constraints, we have not yet evaluated our framework on mixture-of-experts (MoE) architectures like DeepSeek-V3; extending our methodology to these advanced models presents a promising research opportunity. Third, there is potential in integrating our algorithms into the sampling process of large language model reinforcement training, such as reinforcement learning with human feedback (RLHF) and post-training with reinforcement learning, where predefined prompts and varying output lengths play a significant role.

% \kai{We could also mention chunked-prefill, tile-quantization effect. ref. Chunked-prefill paper.}
% \bowen{Mention training process for RLHF.}

\bibliographystyle{ieeetr}
\bibliography{main}

\begin{thebibliography}{10}

\bibitem{achiam2023gpt}
J.~Achiam, S.~Adler, S.~Agarwal, L.~Ahmad, I.~Akkaya, F.~L. Aleman, D.~Almeida, J.~Altenschmidt, S.~Altman, S.~Anadkat, {\em et~al.}, ``Gpt-4 technical report,'' {\em arXiv preprint arXiv:2303.08774}, 2023.

\bibitem{jaech2024openai}
A.~Jaech, A.~Kalai, A.~Lerer, A.~Richardson, A.~El-Kishky, A.~Low, A.~Helyar, A.~Madry, A.~Beutel, A.~Carney, {\em et~al.}, ``Openai o1 system card,'' {\em arXiv preprint arXiv:2412.16720}, 2024.

\bibitem{touvron2023llama}
H.~Touvron, T.~Lavril, G.~Izacard, X.~Martinet, M.-A. Lachaux, T.~Lacroix, B.~Rozi{\`e}re, N.~Goyal, E.~Hambro, F.~Azhar, {\em et~al.}, ``Llama: Open and efficient foundation language models,'' {\em arXiv preprint arXiv:2302.13971}, 2023.

\bibitem{dubey2024llama}
A.~Dubey, A.~Jauhri, A.~Pandey, A.~Kadian, A.~Al-Dahle, A.~Letman, A.~Mathur, A.~Schelten, A.~Yang, A.~Fan, {\em et~al.}, ``The llama 3 herd of models,'' {\em arXiv preprint arXiv:2407.21783}, 2024.

\bibitem{zeng2021pangu}
W.~Zeng, X.~Ren, T.~Su, H.~Wang, Y.~Liao, Z.~Wang, X.~Jiang, Z.~Yang, K.~Wang, X.~Zhang, {\em et~al.}, ``Pangu-$\alpha$: Large-scale autoregressive pretrained chinese language models with auto-parallel computation,'' {\em arXiv preprint arXiv:2104.12369}, 2021.

\bibitem{liu2024deepseek}
A.~Liu, B.~Feng, B.~Xue, B.~Wang, B.~Wu, C.~Lu, C.~Zhao, C.~Deng, C.~Zhang, C.~Ruan, {\em et~al.}, ``Deepseek-v3 technical report,'' {\em arXiv preprint arXiv:2412.19437}, 2024.

\bibitem{guo2025deepseek}
D.~Guo, D.~Yang, H.~Zhang, J.~Song, R.~Zhang, R.~Xu, Q.~Zhu, S.~Ma, P.~Wang, X.~Bi, {\em et~al.}, ``Deepseek-r1: Incentivizing reasoning capability in llms via reinforcement learning,'' {\em arXiv preprint arXiv:2501.12948}, 2025.

\bibitem{chatgpt}
OpenAI, ``Chatgpt.'' \url{https://chatgpt.com/}, 2025.

\bibitem{githubcopilot}
Github, ``Github copilot.'' \url{https://github.com/features/copilot}, 2025.

\bibitem{microsoftcopilot}
Microsoft, ``Microsoft copilot.'' \url{https://copilot.microsoft.com/}, 2025.

\bibitem{leviathan2023fast}
Y.~Leviathan, M.~Kalman, and Y.~Matias, ``Fast inference from transformers via speculative decoding,'' in {\em International Conference on Machine Learning}, pp.~19274--19286, PMLR, 2023.

\bibitem{chen2023accelerating}
C.~Chen, S.~Borgeaud, G.~Irving, J.-B. Lespiau, L.~Sifre, and J.~Jumper, ``Accelerating large language model decoding with speculative sampling,'' {\em arXiv preprint arXiv:2302.01318}, 2023.

\bibitem{dao2022flashattention}
T.~Dao, D.~Y. Fu, S.~Ermon, A.~Rudra, and C.~R{\'e}, ``Flash{A}ttention: Fast and memory-efficient exact attention with {IO}-awareness,'' in {\em Advances in Neural Information Processing Systems (NeurIPS)}, 2022.

\bibitem{kwon2023efficient}
W.~Kwon, Z.~Li, S.~Zhuang, Y.~Sheng, L.~Zheng, C.~H. Yu, J.~Gonzalez, H.~Zhang, and I.~Stoica, ``Efficient memory management for large language model serving with pagedattention,'' in {\em Proceedings of the 29th Symposium on Operating Systems Principles}, pp.~611--626, 2023.

\bibitem{pang2025hybrid}
B.~Pang, K.~Li, R.~She, and F.~Wang, ``Hybrid offline-online scheduling method for large language model inference optimization,'' {\em arXiv preprint arXiv:2502.15763}, 2025.

\bibitem{jaillet2025online}
P.~Jaillet, J.~Jiang, C.~Podimata, and Z.~Zhou, ``Online scheduling for llm inference with kv cache constraints,'' {\em arXiv preprint arXiv:2502.07115}, 2025.

\bibitem{wu2023fast}
B.~Wu, Y.~Zhong, Z.~Zhang, S.~Liu, F.~Liu, Y.~Sun, G.~Huang, X.~Liu, and X.~Jin, ``Fast distributed inference serving for large language models,'' {\em arXiv preprint arXiv:2305.05920}, 2023.

\bibitem{zhong2024distserve}
Y.~Zhong, S.~Liu, J.~Chen, J.~Hu, Y.~Zhu, X.~Liu, X.~Jin, and H.~Zhang, ``$\{$DistServe$\}$: Disaggregating prefill and decoding for goodput-optimized large language model serving,'' in {\em 18th USENIX Symposium on Operating Systems Design and Implementation (OSDI 24)}, pp.~193--210, 2024.

\bibitem{patel2024splitwise}
P.~Patel, E.~Choukse, C.~Zhang, A.~Shah, {\'I}.~Goiri, S.~Maleki, and R.~Bianchini, ``Splitwise: Efficient generative llm inference using phase splitting,'' in {\em 2024 ACM/IEEE 51st Annual International Symposium on Computer Architecture (ISCA)}, pp.~118--132, IEEE, 2024.

\bibitem{yu2022orca}
G.-I. Yu, J.~S. Jeong, G.-W. Kim, S.~Kim, and B.-G. Chun, ``Orca: A distributed serving system for $\{$Transformer-Based$\}$ generative models,'' in {\em 16th USENIX Symposium on Operating Systems Design and Implementation (OSDI 22)}, pp.~521--538, 2022.

\bibitem{agrawal2024sarathi}
A.~Agrawal, N.~Kedia, A.~Panwar, J.~Mohan, N.~Kwatra, B.~S. Gulavani, A.~Tumanov, and R.~Ramjee, ``Taming throughput-latency tradeoff in llm inference with sarathi-serve,'' {\em arXiv preprint arXiv:2403.02310}, 2024.

\end{thebibliography}

\end{document}